\documentclass[aip,reprint,floatfix]{revtex4-2}
\usepackage{amsfonts,amssymb,amsmath,array}

\usepackage{dcolumn}
\usepackage{bm}
\usepackage[final]{graphicx}
\usepackage{graphicx}
\usepackage{epstopdf}
\graphicspath{{./Figures/}} %Setting the graphicspath

\usepackage[usenames]{color}

\usepackage{soul}

\newcommand{\beq}{\begin{equation}}
\newcommand{\eeq}{\end{equation}}
\newcommand{\bea}{\begin{eqnarray}}   
\newcommand{\eea}{\end{eqnarray}}

\begin{document}

\title{Translational and Rotational Temperature Difference in Coexisting Phases of Inertial Active Dumbbells}

\author{Subhasish Chaki}
	\email{subhasischaki@gmail.com}
	\affiliation{
		Institut f{\"u}r Theoretische Physik II: Weiche Materie,
		Heinrich-Heine-Universit{\"a}t D{\"u}sseldorf, Universit{\"a}tsstra{\ss}e 1,
		D-40225 D{\"u}sseldorf, 
		Germany}

\author{Hartmut L{\"o}wen}
	\affiliation{
		Institut f{\"u}r Theoretische Physik II: Weiche Materie,
		Heinrich-Heine-Universit{\"a}t D{\"u}sseldorf, Universit{\"a}tsstra{\ss}e 1,
		D-40225 D{\"u}sseldorf, 
		Germany}

\begin{abstract}
 We investigate the effect of translational and rotational inertia on motility-induced phase separation in underdamped active dumbbells and identify the emergence of four distinct kinetic temperatures across the coexisting phases—unlike in overdamped systems. We find that the dilute, gas-like phase consistently exhibits a higher translational kinetic temperature than the dense, liquid-like phase, with this difference amplified by increasing the rotational inertia. Rotational kinetic temperatures display a similar trend, with the dense phase remaining colder than the dilute phase; however, in this case the temperature difference grows with translational inertia and activity, while becoming practically independent of rotational inertia. This counterintuitive behavior arises from the interplay of activity-driven collisions with both translational and rotational inertia in the coexisting phases. Our results highlight the critical role of translational and rotational inertia in shaping the kinetic temperature landscape of motility-induced phase separation and offer new insights into the nonequilibrium thermodynamics of active matter.
\end{abstract}

\maketitle

\section{Introduction}
\label{sec:introduction}

\noindent Active particles continuously consume energy from their surroundings at the individual level and convert it into directed or persistent motion ~\cite{marchetti2013hydrodynamics,Elgeti2015,bechinger2016active,shaebani2020computational,bowick2022symmetry,te2025exactly,godec2015signal}. As a result, their dynamics fundamentally deviate from those of passive Brownian particles, exhibiting non-equilibrium behavior across a wide range of length scales. Prominent examples include synthetic active colloids ~\cite{buttinoni2013dynamical,vuijk2022active}, motile bacteria ~\cite{xu2023geometrical, aranson2022bacterial}, crawling cells ~\cite{henkes2020dense,thapa2019transient}, and even larger-scale organisms such as fish, birds ~\cite{attanasi2014information}, and insects. A key signature of this non-equilibrium nature is the violation of the fluctuation-dissipation theorem, which manifests in the enhanced long-time diffusion due to persistent motion. Importantly, this persistence not only governs single-particle dynamics but also plays a central role in collective phenomena such as motility-induced phase separation (MIPS) ~\cite{Redner2013Structure,fily2012athermal,palacci2013living,buttinoni2013dynamical,cates2015motility,adorjani2024motility,mcdermott2023characterizing} where, beyond a threshold in activity and density, the system spontaneously separates into a dense phase and a dilute phase in the absence of attractive interactions or alignment.  

 At the microscale, particle inertia is negligible compared to viscous drag from the surrounding solvent. As a result, the instantaneous velocity of an individual particle (distinct from the coarse-grained velocity arising from self-propulsion) relaxes rapidly, leading to overdamped dynamics. In such systems, MIPS closely resembles equilibrium liquid-gas coexistence, with equal kinetic temperatures in both phases. In contrast, for macroscopic active particles where inertia cannot be neglected ~\cite{lowen2020inertial}, the behavior of MIPS is qualitatively altered. In the dense phase, inertial particles repeatedly bounce back upon collisions, a mechanism that generates hotter the gas-like phase compared to the liquid-like phase and can significantly suppress phase separation ~\cite{mandal2019motility,hecht2022active,feng2025theory}.  Interestingly, Caprini $et$ $al.$ demonstrated that rotational inertia, unlike translational inertia, favors MIPS by enhancing the effective persistence time of particle trajectories ~\cite{caprini2022role}. In another study, MIPS has been observed in systems of soft self-propelled disks in the overdamped limit, whereas inertial disks exhibit MIPS only in the hard particle limit ~\cite{de2022motility,feng2025theory}. More recently, Hecht $et.$ $al.$ investigated mixtures of overdamped active and inertial passive Brownian particles ~\cite{hecht2024motility}, finding that the dense, liquid-like phase can be either colder or hotter than the surrounding dilute, gas-like phase—highlighting the complex interplay between collective behavior and inertia.
 
 While most studies on motility-induced phase separation (MIPS) focus on active particles with spherical shapes, natural and artificial active matter systems are predominantly composed of anisotropic particles. Notable examples of anisotropic active matter include rod-shaped bacteria ~\cite{ursell2014rod}, chemically powered nanorods ~\cite{paxton2004catalytic}, and vibrated granular rods ~\cite{kudrolli2008}. Unlike active disks, active rods can slide past one another, and thus typically do not exhibit MIPS ~\cite{duman2018collective,abkenar2013collective}. However, Suma et al. investigated the phase behavior of a suspension of overdamped active rigid dumbbells and demonstrated that the region of phase space where phase separation occurs is significantly broader than that observed for overdamped spherical active particles ~\cite{cugliandolo2017phase,suma2014motility,petrelli2018active}. 
 
 The impact of inertia on the shape and dynamics of macroscopic active systems remains an open and largely unexplored question. To investigate the role of inertia in anisotropic systems, we study a collection of active rigid dumbbells where the active force is applied along their main axis. We focus on uncovering a related and previously overlooked phenomenon: the emergence of a kinetic temperature gradient between coexisting phases in inertial anisotropic active matter. In particular, we examine how inertia influences the rotational kinetic temperature in these phases —a feature absents in spherical particles, where rotational dynamics are decoupled from activity ~\cite{harth2018free,trittel2017mechanical,sprenger2023dynamics,valecha2025active}.  We reveal a mechanism by which rotational inertia gives rise to a kinetic temperature difference between coexisting phases. These findings offer new insights and potential strategies for controlling temperature gradients in the rapidly growing field of macroscopic active matter ~\cite{hecht2024define,schiltz2023kinetic,akintunde2025single}. 

The paper is structured as follows: After introducing the model in Sec.~\ref{sec:model}, we numerically study the translational and rotational temperature difference in dense and dilute phases of inertial active dumbbells in Sec.~\ref{sec:numerics}. Finally, we present a discussion in the conclusive section.

\section{Model}
\label{sec:model}

\begin{figure}[h!]
	\centering
	\includegraphics[width=0.7\linewidth]{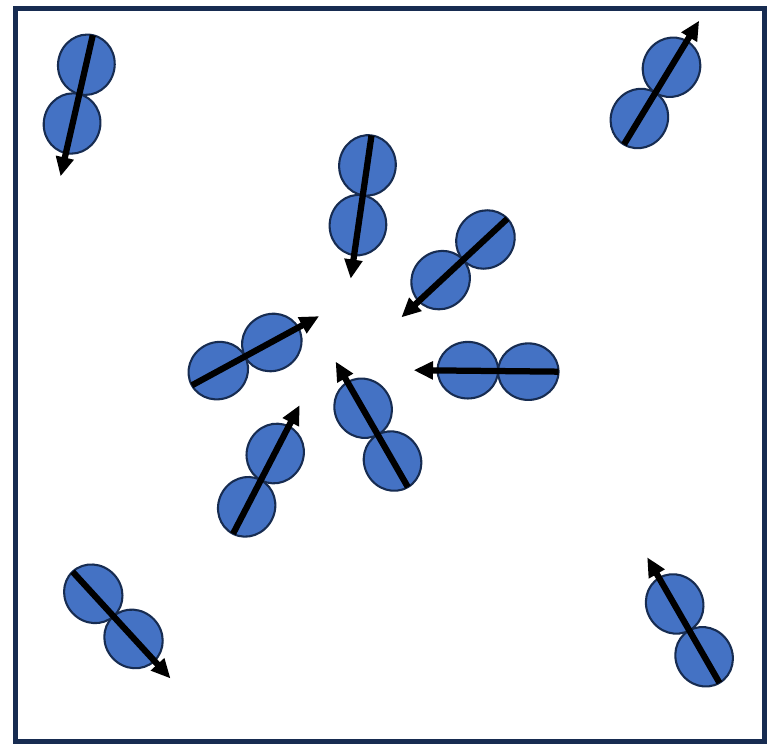}
	\caption{\small Schematic illustration of active dumbbells. Induced by a spontaneous fluctuation, a subset of dumbbells moves toward a common center, leading to the formation of a cluster. Arrows indicate the instantaneous self-propulsion directions.}\label{fig:schematic}
	\label{fig:fig_1}
\end{figure}

\noindent We consider a two-dimensional system of $N$ self-propelled rigid dumbbells confined in a square box of side length $L$ with periodic boundary conditions. Fig.~\ref{fig:fig_1} represents an illustration of the system. Each rigid dumbbell consists of two spherical beads of diameter $\sigma_d$ and mass $m$ and the center-to-center distance between the beads is fixed at $\sigma_d$. The position $\mathbf{r}_i^{(j)}(t)$ of the $j$-th bead of $i$-th dumbbell evolves in time according to the underdamped Langevin equation

\begin{align}    
	m\ddot{\boldsymbol{r}}_i^{(j)} &= -\gamma \dot{\boldsymbol{r}}_i^{(j)} + \boldsymbol{\nabla}_i^{(j)} U  + \boldsymbol{\xi}_i^{(j)} + \boldsymbol{F}_i^{\mathrm{act}}\,,
	\label{eq:eulero_pos1}
\end{align}

\noindent where $i = 1 ,. . ., N$, $j=1,2$, $\gamma$ is the viscous damping coefficient, $U$ denotes the total interaction potential energy, and $\mathbf{F}_i^{\mathrm{act}}$ represents the self-propulsion force. Beads belonging to different dumbbells interact via a generalized Mie potential, defined as $U(r) = \{4\epsilon \left[ \left( \frac{\sigma}{r} \right)^{2n} - \left( \frac{\sigma}{r} \right)^n \right] + \epsilon\}\theta\left(2^{1/n}\sigma-r\right)$ ~\cite{mie1903kinetischen}. The Mie potential is truncated at its minimum, located at $r=2^{1/n}\sigma$, so that the interaction remains purely repulsive. To approximate the hard-disk limit, we choose a large exponent $n=32$.  The cutoff distance is set to $2^{1/n}\sigma=\sigma_d$ to avoid any discontinuities in the force. The stochastic term $\boldsymbol{\xi}_i^{(j)}$ is a Gaussian white noise with zero mean and covariance determined by the fluctuation–dissipation theorem:

\begin{align} 
\langle \boldsymbol{\xi}_{i}^{(j)}(t) \rangle = 0, \quad \langle \boldsymbol{\xi}_{i}^{(j)}(t) \cdot \boldsymbol{\xi}_{i^{\prime}}^{(j^{\prime})}(t') \rangle = 2 \gamma k_{\mathrm{B}} T \, \delta_{ii^{\prime}} \delta_{jj^{\prime}} \, \delta(t - t')
\label{eq:noise}
\end{align}

\noindent where $k_B$ is the Boltzmann constant and $T$ is the ambient temperature. The self-propulsion force acting on each dumbbell is modeled as $\mathbf{F}_i^{\mathrm{act}} = f_a \frac{\left( \mathbf{r}_{i}^{(2)} - \mathbf{r}_{i}^{(1)} \right)}{\left| \mathbf{r}_{i}^{(2)}- \mathbf{r}_{i}^{(1)}\right|}$ where $f_a$ denotes the self-propulsion strength, and $\mathbf{r}_{i}^{(2)} - \mathbf{r}_{i}^{(1)}$ defines the bond vector connecting the $j=2$ and $j=1$ monomers of $i$-th dumbbell. 
\\
\\
\noindent  The dynamics of a rigid dumbbell can be decomposed into translational motion of its center of mass and rotational motion about the center of mass. The net force and torque acting on the dumbbell are obtained by summing the individual forces and torques acting on its constituent disks. The angular velocity $\omega_i$ of the $i$th rigid dumbbell is identical for both $j=2$ and $j=1$ disks and is updated according to
 
\begin{align}    
	I\dot{\boldsymbol{\omega}}_i= \sum_{j=1}^{2} \left[(\boldsymbol{r}_i^{(j)}-\boldsymbol{R}_{\mathrm{com}})\times \boldsymbol{\nabla}_i^j U\right] -\frac{\gamma \sigma_d^2}{2} \boldsymbol{\omega}_i  + \sqrt{k_BT
	\gamma\sigma_d^2} \boldsymbol{\eta}_{ri} 
	\label{eq:orientation}
\end{align}

\noindent where  $I$ is the moment of inertia and $\mathbf{R}_{\mathrm{com}}$ denote the position of the center of mass of each dumbbell. The center of mass of the $i$th dumbbell obeys

\begin{align}    
	2m\dot{\boldsymbol{V}}_{\mathrm{com,i}}=  -2\gamma {\boldsymbol{V}}_{\mathrm{com,i}}+\sum_{j=1}^{2} \boldsymbol{\nabla}_i^{(j)} U +2\boldsymbol{F}_i^{\mathrm{act}} +\sqrt{4k_BT\gamma}\boldsymbol{\eta}_{ti}  
	\label{eq:comvelocity}
\end{align}

\noindent where $\boldsymbol{V}_{\mathrm{com,i}}$ denote the velocity of the center of mass of $i$th dumbbell. $\boldsymbol{\eta}_{ti}$ and $\boldsymbol{\eta}_{ri} $ are Gaussian random forces and torques on the center of mass of the $i$th dumbbell with zero mean and unit variances respectively.  The translational velocity of each disk is then given by $\dot{\boldsymbol{r}}_i^{(j)}=\boldsymbol{V_{\mathrm{com,i}}} + \boldsymbol{\omega}_i \times (\boldsymbol{r}_i^{(j)} - \boldsymbol{R}_{\mathrm{com}}) $. All physical quantities are expressed in reduced units based on the characteristic scales of the system: mass $m$, length  $\sigma_d$, and the energy scale $\epsilon$. Accordingly, the unit of time is defined as $\tau=\sqrt{\frac{m\sigma_d^2}{\epsilon}}$. Throughout this work, we set the thermal energy to $k_B T=0.01 \epsilon$. 
\\
\\
\noindent The behavior of the system is governed by several key dimensionless parameters. The first is the Péclet number, which quantifies the relative strength of active forces compared to thermal fluctuations, and is defined in eq.~\ref{eq:peclet}  as

\begin{align}    
	\mathrm{Pe} = \frac{2\sigma_d f_a}{k_{\mathrm{B}} T} 
	\label{eq:peclet}
\end{align}
  
\noindent  The second dimensionless parameter $\Gamma$ in eq.~\ref{eq:Gamma} characterizes the importance of inertial effects and interpreted as the ratio of two timescales: the momentum relaxation time $\tau_m=\frac{m}{\gamma}$, and the time $\tau_a=\frac{\sigma_d}{f_a/\gamma}$ it takes for a particle to travel its own diameter under self-propulsion. 
 
 \begin{align}    
 	\Gamma = \frac{m f_a}{\gamma^2 \sigma_d}
 	\label{eq:Gamma}
 \end{align} 
 
\noindent  The third parameter is the area fraction defined in eq.~\ref{eq:phi} as 

 \begin{align}    
	\phi = \frac{N \pi \sigma_d^2}{2L^2}
	\label{eq:phi}
\end{align} 

\noindent The simulations are performed using a custom-modified version of the LAMMPS software ~\cite{thompson2022lammps}.

\section{Results}
\label{sec:numerics}

\begin{figure*}[!t]
\centering
\includegraphics[width=1\linewidth,keepaspectratio]{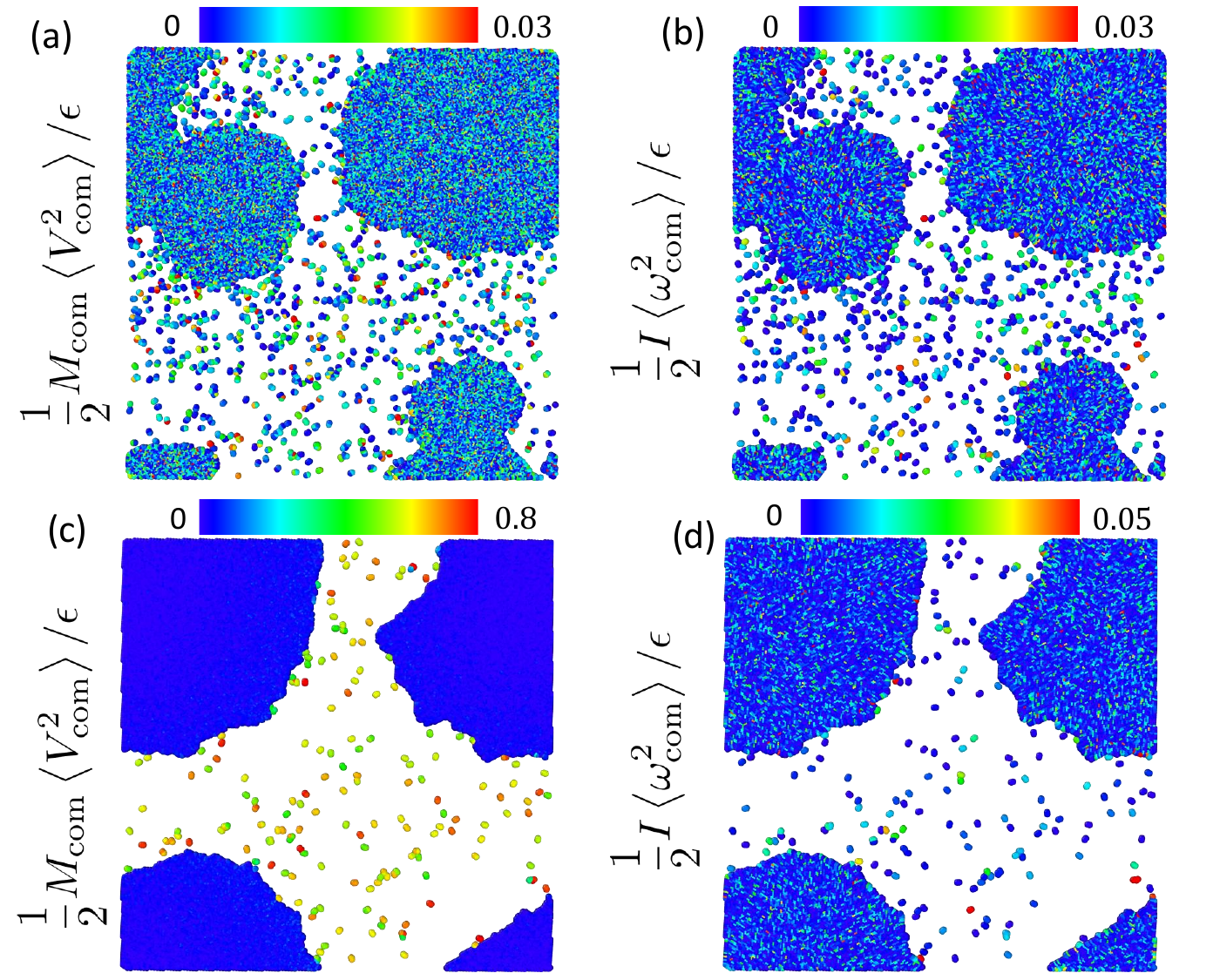}
\caption{
Panels (a), (b), (c), and (d) show snapshots from our simulations in the steady state. Panels (a) and (c) are colored according to the translational kinetic energy, $\frac{1}{2}M_\mathrm{com}\left<V_\mathrm{com}^2\right>$ of individual dumbbells, while panels (b) and (d) are colored according to the rotational kinetic energy, $\frac{1}{2}I\left<\omega_\mathrm{com}^2\right>$, both expressed in units of $\epsilon$. All simulations are performed at area fraction $\phi=0.4$ with $N=16129$ dumbbells. The parameters for (a) and (b) are $\text{Pe}=100$, $\Gamma=0.01$ and $I/m\sigma_d^2=0.5$ and the same for (c) and (d) are $\text{Pe}=100$, $\Gamma=5.06$ and $I/m\sigma_d^2=1$. 
    }
    \label{fig:fig_2}
\end{figure*}

\begin{figure*}[!t]
	\centering
	\includegraphics[width=1\linewidth,keepaspectratio]{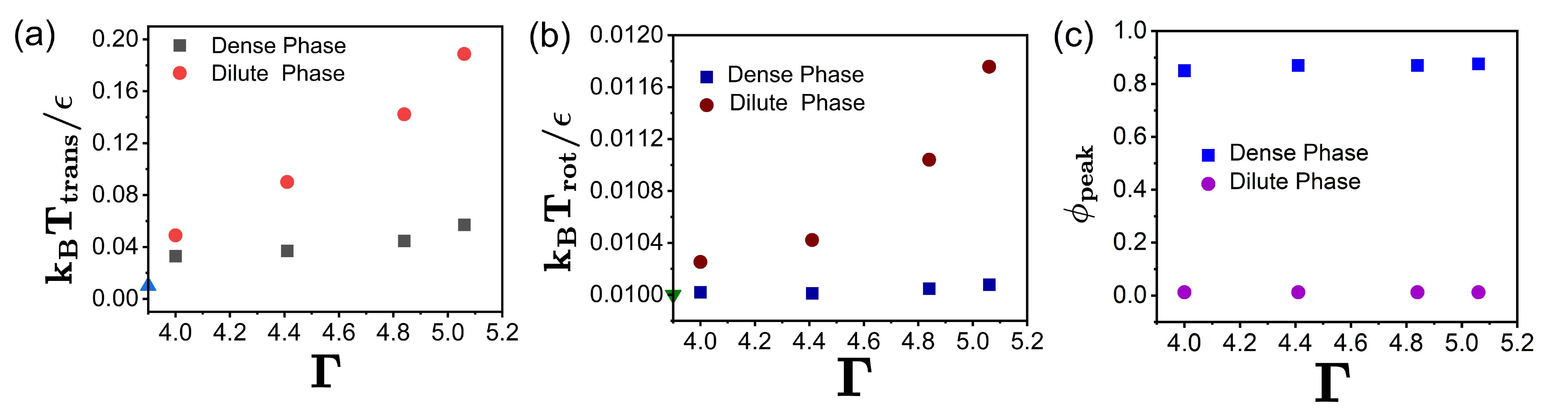}
	\caption{
		Panels (a), (b), and (c) show the translational kinetic temperature, rotational kinetic temperature, and the local packing fraction peak value in the coexisting dense and dilute phases of active dumbbells as functions of the dimensionless translational inertia $\Gamma$ for $\text{Pe}=100$, $I/m\sigma_d^2=0.5$ and $\phi=0.4$. The ambient temperature, $T=0.01$ is indicated by the upright triangle on the $T_{\mathrm{trans}}$-axis in (a) and the inverted triangle on the $T_{\mathrm{rot}}$-axis in (b).
	}
	\label{fig:fig_3}
\end{figure*}

\noindent When active forces dominate over thermal fluctuations, rigid dumbbells with orientations that, on average, point toward a common center exert opposing forces on each other (Fig.~\ref{fig:fig_1}). This leads to the formation of persistent clusters, which continuously grow through coarsening, ultimately resulting in a phase-separated state. In this work, we investigate the translational and rotational kinetic temperatures of the dumbbell system within the phase-separated regime, defined respectively as $\frac{1}{2}M_\mathrm{com}\left<V_\mathrm{com}^2\right>$ and $\frac{1}{2}I\left<\omega_\mathrm{com}^2\right>$. Following the law of energy equipartition, we define the translational and rotational kinetic temperatures in eqs.~\ref{eq:T_trans} and ~\ref{eq:T_rot} respectively as 

\begin{align}    
	T_{\mathrm{trans}}=\frac{1}{2}M_\mathrm{com}\left<V_\mathrm{com}^2\right>/k_B
	\label{eq:T_trans}
\end{align} 

\begin{align}    
    T_{\mathrm{rot}}=I\left<\omega_\mathrm{com}^2\right>/k_B
	\label{eq:T_rot}
\end{align} 

\noindent  To ensure overdamped dynamics, we set the friction coefficient to $\gamma=10$, such that the characteristic time scales associated with translational and rotational inertia remain much smaller than the persistence time of an individual dumbbell, $\tau_p=\frac{\gamma \sigma_d^2}{2k_B T}$. For $Pe=100$, $\Gamma=0.01$, $I/m\sigma_d^2=0.5$ and $\phi=0.4$, the active dumbbell system undergoes a phase separation reminiscent of gas-liquid coexistence and we find that the translational and rotational kinetic temperatures of the dumbbells remain practically  identical in both coexisting phases (Figs.~\ref{fig:fig_2}a and \ref{fig:fig_2}b). Next, we decrease $\gamma$ to increase $\Gamma$, making the dynamics of the dumbbells underdamped. We still observe the formation of small clusters, followed by coarsening, which ultimately leads to complete phase separation. It is noteworthy that high translational inertia typically favors the homogeneous phase when active particles interact via soft repulsive potentials ~\cite{mandal2019motility}, as active particles tend to pass through one another with effectively no collisions, suppressing clustering. In contrast, in our system, the interaction between active dumbbells is more hard-sphere-like, which facilitates blocking of the active dumbbells and promote MIPS even at high translational inertia. In previous studies ~\cite{mandal2019motility}, the translational kinetic energy of active disks in the dense phase was found to be lower than in the dilute phase. A similar trend for $\frac{1}{2}M_\mathrm{com}\left<V_\mathrm{com}^2\right>$ is observed in our simulations of active dumbbells at $Pe=100$, $\Gamma=5.06$, $I/m\sigma_d^2=1.0$ and $\phi=0.4$, where the dense phase appears translationally colder than the dilute phase (Fig.~\ref{fig:fig_2}c). However, in this case, we do not observe any significant difference in the rotational kinetic energy, $\frac{1}{2}I\left<\omega_\mathrm{com}^2\right>$ between the coexisting phases in Fig.~\ref{fig:fig_2}d, despite the fact that the interactions responsible for rotational motion are influenced by activity. This indicates that, in anisotropic underdamped active systems, energy dissipation mechanisms for translation and rotation can differ markedly between the coexisting phases of MIPS.

\begin{figure*}[!t]
	\centering
	\includegraphics[width=1\linewidth,keepaspectratio]{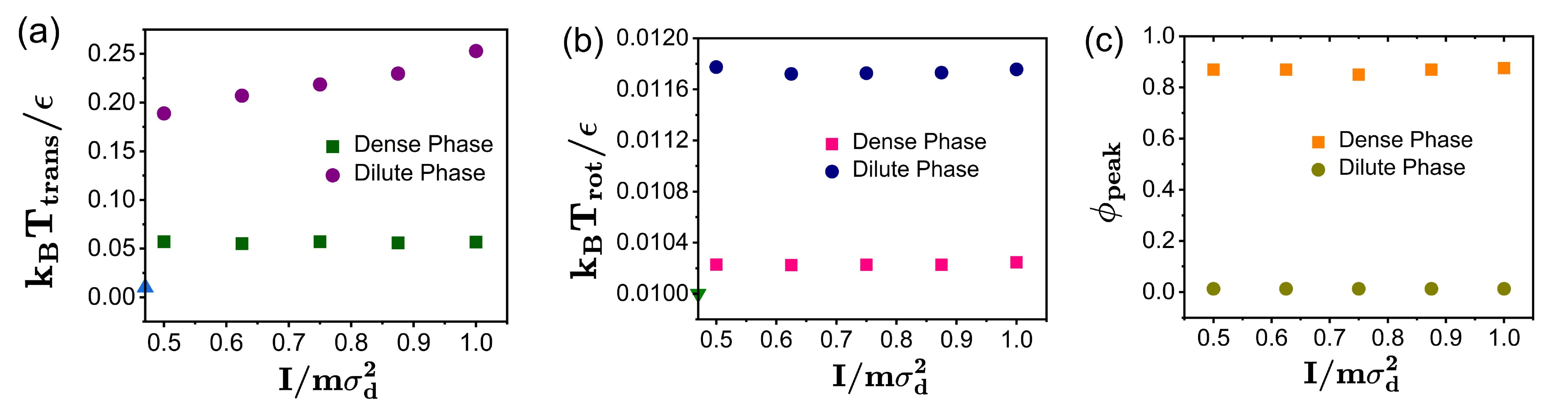}
	\caption{
	Panels (a), (b), and (c) show the translational kinetic temperature, rotational kinetic temperature, and the local packing fraction peak value in the coexisting dense and dilute phases of active dumbbells as functions of the dimensionless rotational inertia $I/m\sigma_d^2$ for $\text{Pe}=100$, $\Gamma=5.06$ and $\phi=0.4$. The ambient temperature, $T=0.01$ is indicated by the upright triangle on the $T_{\mathrm{trans}}$-axis in (a) and the inverted triangle on the $T_{\mathrm{rot}}$-axis in (b).
	}
	\label{fig:fig_4}
\end{figure*}

\begin{figure*}[!t]
	\centering
	\includegraphics[width=0.8\linewidth,keepaspectratio]{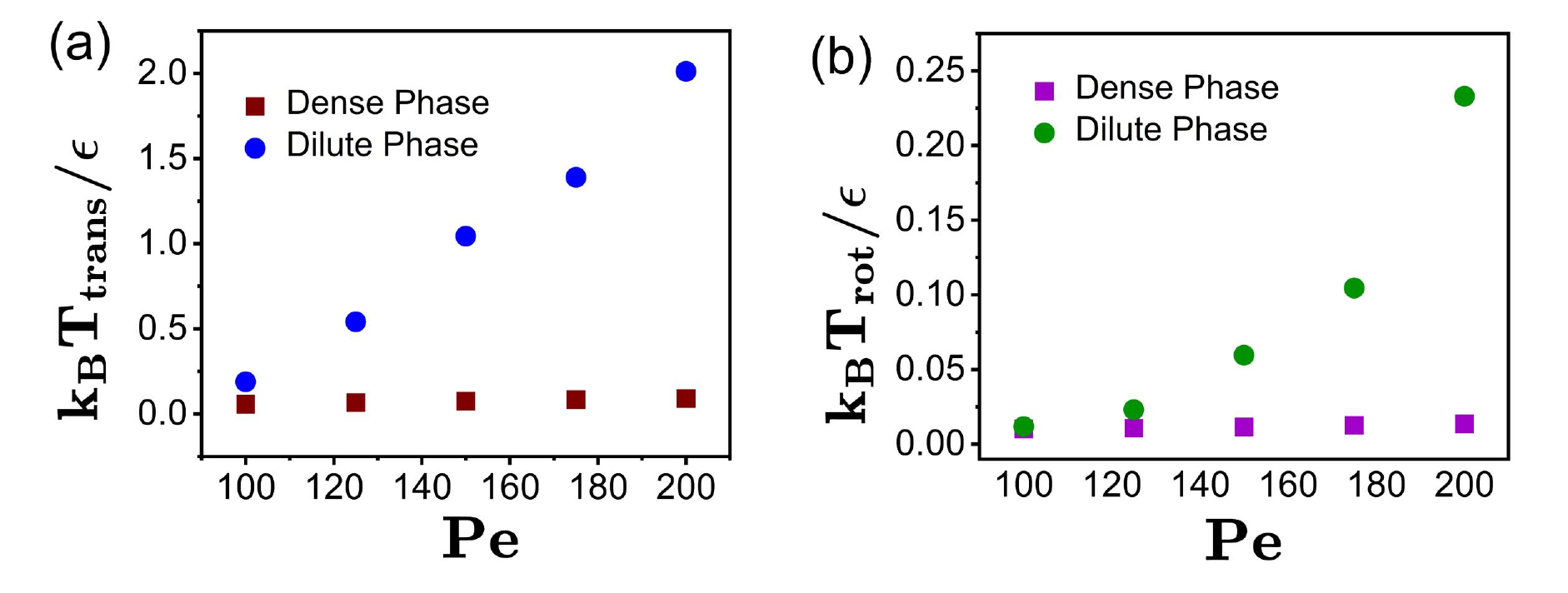}
	\caption{
		Panels (a) and (b) show the translational kinetic temperature and rotational kinetic temperature in the coexisting dense and dilute phases of active dumbbells as functions of the dimensionless Peclet number $\mathrm{Pe}$ for $\Gamma=5.06$, $I/m\sigma_d^2=0.5$ and $\phi=0.4$.
	}
	\label{fig:fig_5}
\end{figure*}

\noindent We choose $\gamma=10\tau^{-1}$ as the reference system, where the dynamics are overdamped and both the translational and rotational kinetic temperatures of the coexisting phases are equal to 0.01, corresponding to the ambient temperature of our simulations as shown in upward-pointing triangle in Fig.~\ref{fig:fig_3}a and the downward-pointing triangle in Fig.~\ref{fig:fig_3}b. To investigate the role of translational inertia, we vary the friction coefficient $\gamma$, noting that the $\Gamma$ scales as $\gamma^2$, while keeping the moment of inertia $I$ and active force $f_a$ fixed. A decrease in $\gamma$ increases the inertial time scale, eventually making it comparable to the persistence time scale, $\tau_p$. As a result, the instantaneous particle velocity becomes governed by both the active force and inertia. We observe that the translational kinetic temperature $T_{\mathrm{trans}}$ in the dilute phase increases substantially with increasing $\Gamma$, whereas the increase is much less pronounced in the dense phase (Fig.~\ref{fig:fig_3}a). In contrast, the difference in rotational kinetic temperature, $T_{\mathrm{rot}}$ between the coexisting phases also grows with $\Gamma$ (Fig.~\ref{fig:fig_3}b), but the difference is much smaller than the corresponding difference in $T_{\mathrm{trans}}$. These observations motivate us to analyze the local packing fraction of the system. We sampled the local packing fraction as follows. First, the system is divided into square cells. For each cell, we compute a coarse-grained local density by averaging the particle density within a circular region of radius $20 \sigma_d$ centered at the cell. From the resulting set of local values, we constructed the corresponding probability distribution function. The dense and dilute phases are identified by distinct peaks in the resulting bimodal distribution of local packing fractions, as is characteristic of MIPS. However, we find that the density difference between the coexisting phases remains nearly unchanged with varying $\Gamma$ (Fig.~\ref{fig:fig_3}c). These results demonstrate that an increase in translational inertia exerts a stronger influence on the particles’ instantaneous velocities than on their local structural organization. Consequently, the translational kinetic temperature difference between the dense and dilute phases increases monotonically. In the dilute phase, infrequent collisions enable dumbbells to sustain relatively high translational velocities under reduced friction. In contrast, within the dense phase, restricted available space and frequent collisions lead to pronounced dissipation of translational kinetic energy, rendering the dense droplet translationally cold. The response of the rotational degrees of freedom exhibits a distinct behavior. In the dilute phase, rotational motion remains largely unaffected by decreasing friction due to the absence of interparticle interactions. In the dense phase, although activity enhances interactions, rotational motion involves minimal spatial rearrangement. As a result, translational inertia has only a weak influence on the rotational kinetic temperature difference between the two phases.
\\
\\
To gain insight into the role of rotational inertia on kinetic temperature differences, we systematically vary the rotational inertia $I$, while keeping the translational inertia parameters—mass $m$, bead separation $\sigma_d$, active force magnitude $f_a$, and friction coefficient $\gamma$—fixed. As shown in Fig.~\ref{fig:fig_4}a, the translational kinetic temperature $T_{\mathrm{trans}}$ in the dilute phase increases significantly with increasing rotational inertia $I/m\sigma_d^2$, while the dense phase remains relatively unaffected. This indicates that the translational kinetic temperature difference between the coexisting phases is more strongly influenced by rotational inertia, $I$ than by translational inertia $\Gamma$. This behavior can be attributed to an increase in the persistence time of active dumbbells with increasing rotational inertia. The active dumbbells exhibit more persistent and directed motion in the dilute phase, enhancing their translational kinetic energy. 
\\
\\
In contrast, in the dense phase, the dumbbells undergo frequent collisions that dissipate translational energy, thereby limiting the translational kinetic temperature increase as compared to the dilute phase. Interestingly, the rotational kinetic temperature exhibits the different trend. With increasing rotational inertia $I$, the rotational kinetic temperature difference in the coexisting phases becomes practically independent of the rotational inertia $I$ (Fig.~\ref{fig:fig_4}b). While frequent dumbbell collisions generate activity-induced torques that enhance rotational motion, this effect is counteracted by rotational inertia, which suppresses it, resulting in a rotational kinetic temperature that is independent of $I$. Similar to (Fig.~\ref{fig:fig_3}c), the density difference between the coexisting phases remains nearly constant with $I$ (Fig.~\ref{fig:fig_4}c). Thus, The active force contributes to rotational motion predominantly through interparticle interactions, emphasizing the crucial role of collisions in transferring activity to rotational degrees of freedom.
\\
\\
Since the observed temperature differences are induced by activity, we now investigate the effect of the self-propulsion speed of the active dumbbells by increasing the Peclet number, $Pe$, while keeping both the translational and rotational inertia constant. We find that the kinetic temperature difference between the translational and rotational degrees of freedom increases with increasing $Pe$ (Fig.~\ref{fig:fig_5}a and ~\ref{fig:fig_5}b). For translational motion, the rise in kinetic temperature with $Pe$ is more pronounced in the dilute phase compared to the dense phase. This is because, in the dense phase, frequent collisions dissipate the energy input from self-propulsion, thereby limiting the increase in kinetic temperature. In contrast, the rotational temperature in the dense phase remains elevated above the ambient temperature due to activity-induced interactions. However, it is only weakly affected by increasing $Pe$, as rotational motion is significantly hindered by crowding effects from neighboring dumbbells. In the dilute phase, although collisions are infrequent, the increase in self-propulsion speed not only reduces the time between successive collisions but also enhances the magnitude of energy transfer to the rotational degrees of freedom, resulting in a $Pe$-dependent increase in rotational kinetic energy. Overall, the dominant contribution to the kinetic temperature differences in both translational and rotational motion arises from the enhanced mobility in the dilute phase relative to the dense phase.

\section{Conclusion}
\label{sec:conclusion}

\noindent In this work, we have investigated the dynamics of inertial active dumbbell systems in the regime of motility-induced phase separation (MIPS), focusing on the kinetic temperature differences between the coexisting dense and dilute phases. As a result, the equipartition of translational and rotational temperature which is valid in equilibrium, is strongly broken by activity. In the phase-separated state there are four different kinetic temperatures which all differ from each other, namely a translational temperature in the dilute and dense phase, and a rotational temperature in the dense and dilute phase. 
\\
\\
 \noindent In particular we have studied here the role of rotational inertia, independent of translational inertia, to probe the effect of activity-induced interactions on kinetic temperature imbalances of the coexisting phases. We find that the system maintains persistent translational and rotational kinetic temperature differences between the coexisting phases during MIPS. Notably, both components of kinetic temperature in the dense and dilute phases exceed the ambient temperature and increase strongly with the self-propulsion speed. This confirms a continuous energy input from the active forces and the absence of thermal equilibrium. Our results reveal that the translational kinetic temperature difference is more sensitive to changes in translational inertia than its rotational counterpart since the rotational motion is strongly influenced by inter-particle collisions than the translational motion. Conversely, increasing rotational inertia significantly impacts the translational kinetic temperature difference by enhancing the persistence of active motion. However, the  rotational kinetic temperature difference remains unaffected with increasing the rotational inertia to maintain the distinct energy-equipartition for rotational degrees of freedom in coexisting phases. These findings highlight the complex kinetic temperature behavior in anisotropic active systems and underscore the importance of particle shape and rotational degrees of freedom in nonequilibrium phase behavior.
 \\
 \\
 \noindent  Our results provide concrete predictions for experimental systems, such as vibrated granular particles, where both translational and rotational inertia can be systematically tuned ~\cite{volfson2004anisotropy,gonzalez2017clustering,yadav2013effect,narayan2007long,scholz2018inertial,caprini2024emergent,deseigne2010collective}. On the theoretical front, a more detailed exploration of the phase diagram and the influence of interaction stiffness on structure and dynamics would provide further insight into inertial MIPS in anisotropic active matter.

\section{Acknowledgments} 
%LC acknowledges support from 
\noindent 
SC acknowledges support from the Alexander Von Humboldt foundation. HL acknowledges support by the Deutsche Forschungsgemeinschaft (DFG) through the SPP 2265, under grant numbers LO 418/25.

\section*{AUTHOR DECLARATIONS}
\noindent
Conflict of Interest: The authors have no conflicts of interest to disclose.

\section*{DATA AVAILABILITY}
\noindent
The data that support the findings of this study are available from the corresponding author upon reasonable request.

%\end{widetext}

%\bibliographystyle{mdpi}
%\bibliographystyle{rsc} %the RSC's .bst file
\bibliographystyle{rsc} %the RSC's .bst file

\bibliography{bib.bib}

@article{caprini2024emergent,
  title={Emergent memory from tapping collisions in active granular matter},
  author={Caprini, Lorenzo and Ldov, Anton and Gupta, Rahul Kumar and Ellenberg, Hendrik and Wittmann, Ren{\'e} and L{\"o}wen, Hartmut and Scholz, Christian},
  journal={Commun. Phys.},
  doi = {10.1038/s42005-024-01540-w},
  volume={7},
  number={1},
  pages={52},
  year={2024},
  publisher={Nature Publishing Group UK London}
}

@article{scholz2018inertial,
  title={Inertial delay of self-propelled particles},
  author={Scholz, Christian and Jahanshahi, Soudeh and Ldov, Anton and L{\"o}wen, Hartmut},
  journal={Nat. Commun.},
doi = {10.1038/s41467-018-07596-x},
  volume={9},
  number={1},
  pages={5156},
  year={2018},
  publisher={Nature Publishing Group UK London}
}

@article{deseigne2010collective,
  title={Collective motion of vibrated polar disks},
  author={Deseigne, Julien and Dauchot, Olivier and Chat{\'e}, Hugues},
  journal={Phys. Rev. Lett.},
doi = {10.1103/PhysRevLett.105.098001},
  volume={105},
  number={9},
  pages={098001},
  year={2010},
  publisher={APS}
}

@article{paxton2004catalytic,
  title={Catalytic nanomotors: autonomous movement of striped nanorods},
  author={Paxton, Walter F and Kistler, Kevin C and Olmeda, Christine C and Sen, Ayusman and St. Angelo, Sarah K and Cao, Yanyan and Mallouk, Thomas E and Lammert, Paul E and Crespi, Vincent H},
  journal={J.  Am. Chem. Soc.},
  volume={126},
  number={41},
  pages={13424--13431},
  year={2004},
  publisher={ACS Publications}
}

@Article{kudrolli2008,
  author    = {Kudrolli, Arshad and Lumay, Geoffroy and Volfson, Dmitri and Tsimring, Lev S},
  title     = {{Swarming and swirling in self-propelled polar granular rods}},
  journal   = {Phys. Rev. Lett.},
  year      = {2008},
  volume    = {100},
  number    = {5},
  pages     = {058001},
  month     = {feb},
  issn      = {0031-9007},
  doi       = {10.1103/PhysRevLett.100.058001},
  publisher = {APS},
}

@article{marchetti2013hydrodynamics,
  title={Hydrodynamics of soft active matter},
  author={Marchetti, M.C. and Joanny, J.F. and Ramaswamy, S. and Liverpool, T.B. and Prost, J. and Rao, M. and Simha, R. A.},
  journal={Rev. Mod. Phys.},
doi = {10.1103/RevModPhys.85.1143},
  volume={85},
  pages={1143--1189},
  year={2013},
  publisher={APS}
}

@article{duman2018collective,
  title={Collective dynamics of self-propelled semiflexible filaments},
  author={Duman, {\"O}zer and Isele-Holder, Rolf E and Elgeti, Jens and Gompper, Gerhard},
  journal={Soft matter},
  volume={14},
  number={22},
  pages={4483--4494},
  year={2018},
  publisher={Royal Society of Chemistry}
}

@article{abkenar2013collective,
  title={Collective behavior of penetrable self-propelled rods in two dimensions},
  author={Abkenar, Masoud and Marx, Kristian and Auth, Thorsten and Gompper, Gerhard},
  journal={Phys. Rev. E},
  volume={88},
  number={6},
  pages={062314},
  year={2013},
  publisher={APS}
}

@article{ursell2014rod,
  title={Rod-like bacterial shape is maintained by feedback between cell curvature and cytoskeletal localization},
  author={Ursell, Tristan S and Nguyen, Jeffrey and Monds, Russell D and Colavin, Alexandre and Billings, Gabriel and Ouzounov, Nikolay and Gitai, Zemer and Shaevitz, Joshua W and Huang, Kerwyn Casey},
  journal={Proc. Nat. Acad. Sci. USA},
  volume={111},
  number={11},
  pages={E1025--E1034},
  year={2014},
  publisher={National Academy of Sciences}
}

@Article{Elgeti2015,
  author   = {Elgeti, J and Winkler, R G and Gompper, G},
  title    = {{Physics of microswimmers—single particle motion and collective behavior: a review}},
  journal  = {Rep. Prog. Phys.},
  year     = {2015},
doi = {10.1088/0034-4885/78/5/056601},
  volume   = {78},
  number   = {5},
  pages    = {56601}
}

@article{bechinger2016active,
  title={Active particles in complex and crowded environments},
  author={Bechinger, Clemens and Di Leonardo, Roberto and L{\"o}wen, Hartmut and Reichhardt, Charles and Volpe, Giorgio and Volpe, Giovanni},
  journal={Rev. Mod. Phys.},
doi = {10.1103/RevModPhys.88.045006},
  volume={88},
  number={4},
  pages={045006},
  year={2016},
  publisher={APS}
}

@article{shaebani2020computational,
  title={Computational models for active matter},
  author={Shaebani, M Reza and Wysocki, Adam and Winkler, Roland G and Gompper, Gerhard and Rieger, Heiko},
  journal={Nat. Rev. Phys.},
doi = {10.1038/s42254-020-0152-1},
  volume={2},
  pages={181--199},
  year={2020},
  publisher={Nature Publishing Group}
}

@article{lowen2020inertial,
  title={Inertial effects of self-propelled particles: From active Brownian to active Langevin motion},
  author={L{\"o}wen, Hartmut},
  journal={J. Chem. Phys.},
doi = {10.1063/1.5134455},
  volume={152},
  number={4},
  pages={040901},
  year={2020},
  publisher={AIP Publishing LLC}
}

@article{xu2023geometrical,
  title={Geometrical control of interface patterning underlies active matter invasion},
  author={Xu, Haoran and Nejad, Mehrana R and Yeomans, Julia M and Wu, Yilin},
  journal={Proc. Nat. Acad. Sci.  USA},
  volume={120},
  number={30},
  pages={e2219708120},
  year={2023},
  publisher={National Academy of Sciences}
}

@article{henkes2020dense,
  title={Dense active matter model of motion patterns in confluent cell monolayers},
  author={Henkes, Silke and Kostanjevec, Kaja and Collinson, J Martin and Sknepnek, Rastko and Bertin, Eric},
  journal={Nat. Commun.},
  volume={11},
  number={1},
  pages={1405},
  year={2020},
  publisher={Nature Publishing Group UK London}
}

@article{attanasi2014information,
  title={Information transfer and behavioural inertia in starling flocks},
  author={Attanasi, Alessandro and Cavagna, Andrea and Del Castello, Lorenzo and Giardina, Irene and Grigera, Tomas S and Jeli{\'c}, Asja and Melillo, Stefania and Parisi, Leonardo and Pohl, Oliver and Shen, Edward and others},
  journal={Nat. Phys.},
  volume={10},
  number={9},
  pages={691--696},
  year={2014},
  publisher={Nature Publishing Group UK London}
}

@article{buttinoni2013dynamical,
  title={Dynamical clustering and phase separation in suspensions of self-propelled colloidal particles},
  author={Buttinoni, Ivo and Bialk{\'e}, Julian and K{\"u}mmel, Felix and L{\"o}wen, Hartmut and Bechinger, Clemens and Speck, Thomas},
  journal={Phys. Rev. Lett.},
doi={10.1103/PhysRevLett.110.238301},
  volume={110},
  number={23},
  pages={238301},
  year={2013},
  publisher={APS}
}

@article{palacci2013living, 
  title={Living crystals of light-activated colloidal surfers},
  author={Palacci, Jeremie and Sacanna, Stefano and Steinberg, Asher Preska and Pine, David J and Chaikin, Paul M},
  journal={Science},
doi={10.1126/science.1230020},
  volume={339},
  number={6122},
  pages={936--940},
  year={2013},
  publisher={American Association for the Advancement of Science}
}

@article{caprini2022role,
  title={Role of rotational inertia for collective phenomena in active matter},
  author={Caprini, Lorenzo and Gupta, Rahul Kumar and L{\"o}wen, Hartmut},
  journal={Phys. Chem. Chem. Phys.},
doi={https://doi.org/10.1039/D2CP02940E},
  volume={24},
  number={40},
  pages={24910--24916},
  year={2022},
  publisher={Royal Society of Chemistry}
}

@article{hecht2022active,
  title={Active refrigerators powered by inertia},
  author={Hecht, Lukas and Mandal, Suvendu and L{\"o}wen, Hartmut and Liebchen, Benno},
  journal={Phys. Rev. Lett.},
doi={10.1103/PhysRevLett.129.178001},
  volume={129},
  number={17},
  pages={178001},
  year={2022},
  publisher={APS}
}

@article{mandal2019motility,
  title={Motility-induced temperature difference in coexisting phases},
  author={Mandal, Suvendu and Liebchen, Benno and L{\"o}wen, Hartmut},
  journal={Phys. Rev. Lett.},
doi={10.1103/PhysRevLett.123.228001},
  volume={123},
  number={22},
  pages={228001},
  year={2019},
  publisher={APS}
}

@article{hecht2024motility,
  title={Motility-induced coexistence of a hot liquid and a cold gas},
  author={Hecht, Lukas and Dong, Iris and Liebchen, Benno},
  journal={Nat. Commun.},
doi={10.1038/s41467-024-47533-9},
  volume={15},
  number={1},
  pages={3206},
  year={2024},
  publisher={Nature Publishing Group UK London}
}

@article{de2022motility,
  title={Motility-induced phase separation of self-propelled soft inertial disks},
  author={De Karmakar, Soumen and Ganesh, Rajaraman},
  journal={Soft Matter},
doi={https://doi.org/10.1039/D2SM00772J},
  volume={18},
  number={38},
  pages={7301--7308},
  year={2022},
  publisher={Royal Society of Chemistry}
}

@article{sprenger2023dynamics,
  title={Dynamics of active particles with translational and rotational inertia},
  author={Sprenger, Alexander R and Caprini, Lorenzo and L{\"o}wen, Hartmut and Wittmann, Ren{\'e}},
  journal={J. Phys. Condens. Matter.},
doi={10.1088/1361-648X/accd36},
  volume={35},
  number={30},
  pages={305101},
  year={2023},
  publisher={IOP Publishing}
}

@article{cates2015motility,
  title={Motility-induced phase separation},
  author={Cates, Michael E and Tailleur, Julien},
  journal={Annu. Rev. Condens. Matter Phys.},
  doi={10.1146/annurev-conmatphys-031214-014710},
  volume={6},
  number={1},
  pages={219--244},
  year={2015},
  publisher={Annual Reviews}
}

@article{feng2025theory,
  title={Theory for the anomalous phase behavior of inertial active Brownian particles},
  author={Feng, Jiechao and Omar, Ahmad K},
  journal={Phys. Rev. E},
  volume={111},
  number={4},
  pages={L043402},
  year={2025},
  publisher={APS}
}

@article{Redner2013Structure,
title={Structure and Dynamics of a Phase-separating Active Colloidal Fluid},
  author={Redner, Gabriel S and Hagan, Michael F and Baskaran, Aparna},
 journal={Phys. Rev. Lett},
doi={10.1016/j.bpj.2012.11.3534},
  volume={110},
  number={055701},
  pages={055701},
  year={2013},
  publisher={APS}
}

@article{petrelli2018active,
  title={Active dumbbells: Dynamics and morphology in the coexisting region},
  author={Petrelli, Isabella and Digregorio, Pasquale and Cugliandolo, Leticia F and Gonnella, Giuseppe and Suma, Antonio},
  journal={Euro. Phys. J. E},
  volume={41},
  number={10},
  pages={128},
  year={2018},
  publisher={Springer}
}

@article{cugliandolo2017phase,
  title={Phase coexistence in two-dimensional passive and active dumbbell systems},
  author={Cugliandolo, Leticia F and Digregorio, Pasquale and Gonnella, Giuseppe and Suma, Antonio},
  journal={Phys. Rev. Lett.},
  volume={119},
  number={26},
  pages={268002},
  year={2017},
  publisher={APS}
}

@article{suma2014motility,
  title={Motility-induced phase separation in an active dumbbell fluid},
  author={Suma, Antonio and Gonnella, Giuseppe and Marenduzzo, Davide and Orlandini, Enzo},
  journal={Europhys. Lett.},
  volume={108},
  number={5},
  pages={56004},
  year={2014},
  publisher={IOP Publishing}
}

@article{fily2012athermal,
  title={Athermal phase separation of self-propelled particles with no alignment},
  author={Fily, Yaouen and Marchetti, M Cristina},
  journal={Phys. Rev. Lett.},
  volume={108},
  number={23},
  doi = {10.1103/PhysRevLett.108.235702},
  pages={235702},
  year={2012},
  publisher={APS}
}

@article{bowick2022symmetry,
  title={Symmetry, thermodynamics, and topology in active matter},
  author={Bowick, Mark J and Fakhri, Nikta and Marchetti, M Cristina and Ramaswamy, Sriram},
  journal={Phys. Rev. X},
  volume={12},
  number={1},
  pages={010501},
  year={2022},
  publisher={APS}
}

@article{te2025exactly,
  title={What exactly is' active matter'?},
  author={te Vrugt, Michael and Liebchen, Benno and Cates, Michael E},
  journal={arXiv},
  pages={2507.21621},
  year={2025}
}

@article{thompson2022lammps,
  title={LAMMPS-a flexible simulation tool for particle-based materials modeling at the atomic, meso, and continuum scales},
  author={Thompson, Aidan P and Aktulga, H Metin and Berger, Richard and Bolintineanu, Dan S and Brown, W Michael and Crozier, Paul S and In't Veld, Pieter J and Kohlmeyer, Axel and Moore, Stan G and Nguyen, Trung Dac and others},
  journal={Comput. Phys. Commun.},
  volume={271},
  pages={108171},
  year={2022},
  publisher={Elsevier}
}

@article{mie1903kinetischen,
  title={Zur kinetischen Theorie der einatomigen K{\"o}rper},
  author={Mie, Gustav},
  journal={Ann. Phys.},
  volume={316},
  number={8},
  pages={657--697},
  year={1903},
  publisher={Wiley Online Library}
}

@article{aranson2022bacterial,
  title={Bacterial active matter},
  author={Aranson, Igor S},
  journal={Rep. Prog. Phys.},
  volume={85},
  number={7},
  pages={076601},
  year={2022},
  publisher={IOP Publishing}
}

@article{harth2018free,
  title={Free cooling of a granular gas of rodlike particles in microgravity},
  author={Harth, Kirsten and Trittel, Torsten and Wegner, Sandra and Stannarius, Ralf},
  journal={Phys. Rev. Lett.},
  volume={120},
  number={21},
  pages={214301},
  year={2018},
  publisher={APS}
}

@article{trittel2017mechanical,
  title={Mechanical excitation of rodlike particles by a vibrating plate},
  author={Trittel, Torsten and Harth, Kirsten and Stannarius, Ralf},
  journal={Phys. Rev. E},
  volume={95},
  number={6},
  pages={062904},
  year={2017},
  publisher={APS}
}

@article{volfson2004anisotropy,
  title={Anisotropy-driven dynamics in vibrated granular rods},
  author={Volfson, Dmitri and Kudrolli, Arshad and Tsimring, Lev S},
  journal={Phys. Rev. E},
  volume={70},
  number={5},
  pages={051312},
  year={2004},
  publisher={APS}
}

@article{gonzalez2017clustering,
  title={Clustering in vibrated monolayers of granular rods},
  author={Gonzalez-Pinto, Miguel and Borondo, Florentino and Mart{\'\i}nez-Rat{\'o}n, Yuri and Velasco, Enrique},
  journal={Soft Matter},
  volume={13},
  number={14},
  pages={2571--2582},
  year={2017},
  publisher={Royal Society of Chemistry}
}

@article{yadav2013effect,
  title={Effect of aspect ratio on the development of order in vibrated granular rods},
  author={Yadav, Vikrant},
  journal={Phys. Rev. E},
  volume={88},
  number={88},
  pages={052203},
  year={2013},
  publisher={APS}
}

@article{narayan2007long,
  title={Long-lived giant number fluctuations in a swarming granular nematic},
  author={Narayan, Vijay and Ramaswamy, Sriram and Menon, Narayanan},
  journal={Science},
  volume={317},
  number={5834},
  pages={105--108},
  year={2007},
  publisher={American Association for the Advancement of Science}
}

@article{hecht2024define,
  title={How to define temperature in active systems?},
  author={Hecht, Lukas and Caprini, Lorenzo and L{\"o}wen, Hartmut and Liebchen, Benno},
  journal={J Chem. Phys.},
  volume={161},
  number={22},
  pages={22},
  year={2024},
  publisher={AIP Publishing}
}

@article{godec2015signal,
  title={Signal focusing through active transport},
  author={Godec, Alja{\v{z}} and Metzler, Ralf},
  journal={Phys. Rev. E},
  volume={92},
  number={1},
  pages={010701},
  year={2015},
  publisher={APS}
}

@article{thapa2019transient,
  title={Transient superdiffusion of polydisperse vacuoles in highly motile amoeboid cells},
  author={Thapa, Samudrajit and Lukat, Nils and Selhuber-Unkel, Christine and Cherstvy, Andrey G and Metzler, Ralf},
  journal={J. Chem. Phys.},
  volume={150},
  number={14},
  pages={14},
  year={2019},
  publisher={AIP Publishing}
}

@article{valecha2025active,
  title={Active transport of cargo-carrying and interconnected chiral particles},
  author={Valecha, Bhavesh and Vahid, Hossein and Muzzeddu, Pietro Luigi and Sommer, Jens-Uwe and Sharma, Abhinav},
  journal={Soft Matter},
  volume={21},
  number={17},
  pages={3384--3392},
  year={2025},
  publisher={Royal Society of Chemistry}
}

@article{vuijk2022active,
  title={Active colloidal molecules in activity gradients},
  author={Vuijk, Hidde D and Klempahn, Sophie and Merlitz, Holger and Sommer, Jens-Uwe and Sharma, Abhinav},
  journal={Phys. Rev. E},
  volume={106},
  number={1},
  pages={014617},
  year={2022},
  publisher={APS}
}

@article{schiltz2023kinetic,
  title={Kinetic temperature and pressure of an active Tonks gas},
  author={Schiltz-Rouse, Elijah and Row, Hyeongjoo and Mallory, Stewart A},
  journal={Phys. Rev. E},
  volume={108},
  number={6},
  pages={064601},
  year={2023},
  publisher={APS}
}

@article{akintunde2025single,
  title={Single-file diffusion of active Brownian particles},
  author={Akintunde, Akinlade and Bayati, Parvin and Row, Hyeongjoo and Mallory, Stewart A},
  journal={J. Chem. Phys.},
  volume={162},
  number={16},
  pages={16},
  year={2025},
  publisher={AIP Publishing}
}

@article{adorjani2024motility,
  title={Motility-induced phase separation and frustration in active matter swarmalators},
  author={Adorj{\'a}ni, B and Lib{\'a}l, A and Reichhardt, Charles and Reichhardt, Cynthia Jane Olson},
  journal={Phys. Rev. E},
  volume={109},
  number={2},
  pages={024607},
  year={2024},
  publisher={APS}
}

@article{mcdermott2023characterizing,
  title={Characterizing different motility-induced regimes in active matter with machine learning and noise},
  author={McDermott, Danielle and Reichhardt, Charles and Reichhardt, Cynthia Jane Olson},
  journal={Phys. Rev. E},
  volume={108},
  number={6},
  pages={064613},
  year={2023},
  publisher={APS}
}

%%%%%%%%%%%%%%%%%%%%%%%%%%%%%%%%%%%%%%%%%%
%% optional
%\sampleavailability{Samples of the compounds ...... are available from the authors.}

\end{document}